\documentclass[aps,prl,preprint,superscriptaddress]{revtex4}

\usepackage{graphicx}
\usepackage{amssymb}

\begin{document}


\title{Signatures for flow effects in $\sqrt{s}=200$ GeV
proton-proton collisions.}


\author{T. J. Humanic}
\email[]{humanic@mps.ohio-state.edu}
\affiliation{Department of Physics, The Ohio State University,
Columbus, Ohio, USA}


\date{\today}

\begin{abstract}
A simple model based on relativistic geometry and final-state hadronic rescattering
is used to predict pion source parameters extracted in two-pion femtoscopy studies of
proton-proton collisions at $\sqrt s=200$ GeV. From studying the momentum and
particle multiplicity dependences of these parameters in the context of this model
and assuming a very short hadronization time, flow-like behavior is seen which 
resembles the flow behavior commonly observed in relativistic heavy-ion collisions.
\end{abstract}


\maketitle


In the field of fluid dynamics, where a fluid can be either a liquid or a gas,
fluid flow is generally defined as the transport of a certain amount of fluid 
mass or volume across a surface in a given time interval. This concept of fluid flow 
has been adopted by the relativistic heavy-ion
collision community to describe the behaviors of some observables seen in experiments
which appear to be flow-like in nature \cite{rhic1,rhic2,rhic3,rhic4}. 
The justification for having a fluid dynamics picture of a
relativistic heavy-ion collision is that in these collisions, for example colliding beams 
of $Au+Au$ at 200 GeV per nucleon pair in the center-of-mass frame, thousands of
particles such as partons, e.g. quarks and gluons, 
hadrons, e.g. nucleons and pions, and leptons, 
e.g. electrons and muons, participate and are generated 
out of the vacuum in a violent collision. Thus the concept of characterizing the ``bulk 
properties'' of such collisions, for example the dynamics of the size and shape of the
interaction region, by a relativistic fluid flow where the fluid is made up of
thousands of particles seems reasonable to some approximation. On the other hand,
if one were to try to impose a fluid dynamics picture on proton-proton collisions
at similar energies, e.g. $p+p$ collisions at $\sqrt s=200$ GeV, in which typically
much fewer than 100 particles participate and are produced in the collision, such an
approximation would seem unreasonable due to the paucity of particles available
to compose the ``fluid.''

The goal of the present work is to use a simple model to study one of these bulk
properties in $p+p$ collisions at $\sqrt s=200$ GeV, namely the dynamics of the size
and shape of the interaction region, using the 
method of two-pion femtoscopy \cite{Humanic:2006a,Lisa:2005a}
also known as
Hanbury-Brown-Twiss interferometry (or HBT, which has
also been used to measure the sizes of stars\cite{Humanic:2006a,hbt1}). The HBT observables,
which are radius parameters, have been shown in $Au+Au$ collisions at $\sqrt{s_{NN}}=200$ GeV
to exhibit radial flow-like behavior in their dependences on the pion momenta and the
particle multiplicity of the collision, such that the radius parameters decrease
for increasing pion momenta (higher velocity fluid with smaller correlation length) and 
increase for increasing particle multiplicity 
(higher density fluid producing more expansion)\cite{Adams:2004yc}.
If the model predicts that there are similar dependences
also present in $p+p$ collisions at $\sqrt s=200$ GeV it would help to establish the possibility of 
flow-like effects in these collisions as well as characterize the mechanism by
which these effects are generated.

The model calculations are carried out in four main steps: 
1) simulate $p+p$  collisions using the standard collision-generator code PYTHIA\cite{pythia6.4}, 
2) employ a simple space-time geometry picture for the hadronization of the
PYTHIA-generated hadrons,
3) calculate the effects of final-state rescattering among the hadrons,
and 4) calculate the HBT radius parameters. These steps will now be discussed in more detail.

The $p+p$ collisions were modeled with the PYTHIA code,
version 6.409 \cite{pythia6.4}. The parton distribution functions used were the same as used in Ref. \cite{Humanic:2006ib}. Events were generated
in ``minimum bias'' mode, i.e. setting the low-$p_T$ cutoff for parton-parton collisions to zero and excluding elastic and diffractive collisions. To obtain good statistics for this study $5\times 10^6$
events were generated with $\sqrt{s}= $ 200 GeV. Information saved
from a PYTHIA run for use in the next step of the procedure were the momenta and identities
of the ``direct'' (i.e. redundancies removed) hadrons (all charge states) $\pi$, $K$, $p$, $n$,
$\Lambda$, $\rho$, $\omega$, $\eta$, ${\eta}'$, $\phi$, and $K^*$. These particles were
chosen since they are the most common hadrons produced and thus should have the greatest
effect on the hadronic observables in these calculations.
 

The space-time geometry picture for hadronization from a  $p+p$
collision in the model consists of the emission of a PYTHIA
particle from a thin uniform disk of radius 1 fm in the $x-y$ plane followed by
its hadronization which occurs in the proper time of the particle, $\tau$. The space-time
coordinates at hadronization in the lab frame $(x_h, y_h, z_h, t_h)$ for a particle with momentum
coordinates $(p_x, p_y, p_z)$, energy $E$, rest mass $m_0$, and transverse disk
coordinates $(x_0, y_0)$, which are chosen randomly on the disk,  can then be written as

\begin{eqnarray}
x_h = x_0 + \tau \frac{p_x}{m_0} \\
y_h = y_0 + \tau \frac{p_y}{m_0} \\
z_h = \tau \frac{p_z}{m_0} \\
t_h = \tau \frac{E}{m_0}
\end{eqnarray}

The simplicity of this geometric picture is now clear: it is just an expression of causality with the
assumption that all particles hadronize with the same proper time, $\tau$. A similar hadronization
picture has been applied to $e^+-e^-$ collisions\cite{csorgo}
and Tevatron $p+p$ collisions\cite{Humanic:2006ib}.
For all results presented in this work,  $\tau$ will be set to 0.1 fm/c to be consistent with the results
found in the Tevatron study which had reasonable agreement with measurements.

The hadronic rescattering calculational method used is similar to that
employed in previous studies \cite{Humanic:2006a,Humanic:2006ib}.
Rescattering is simulated with a semi-classical Monte Carlo
calculation which assumes strong binary collisions between hadrons.
Relativistic kinematics is used throughout. The hadrons considered in the
calculation are the most common ones: pions, kaons,
nucleons and lambdas ($\pi$, K,
N, and $\Lambda$), and the $\rho$, $\omega$, $\eta$, ${\eta}'$,
$\phi$, $\Delta$, and $K^*$ resonances.
Although $\Delta$ formation was included in the rescattering process
due to its large formation cross section in 
meson-baryon and baryon-baryon rescattering,
they were
not input directly from PYTHIA since they decay quickly and relatively few
are initially produced in these collisions.
For simplicity, the
calculation is isospin averaged (e.g. no distinction is made among a
$\pi^{+}$, $\pi^0$, and $\pi^{-}$).
Starting from the
initial stage ($t=0$ fm/c), the positions of all particles in each event are
allowed to evolve in time in small time steps ($\Delta t=0.1$ fm/c)
according to their initial momenta. At each time step each particle
is checked to see a) if it has hadronized ($t>t_h$, where $t_h$ is given in
Eq. (4)), b) if it
decays, and c) if it is sufficiently close to another particle to
scatter with it. Isospin-averaged s-wave and p-wave cross sections
for meson scattering are obtained from Prakash et al.\cite{Prakash:1993a}
and other cross sections are estimated from fits to hadron scattering data
in the Review of Particle Physics\cite{pdg}. Both elastic and inelastic collisions are
included. The calculation is carried out to 50 fm/c which
allows ample time for the rescattering to finish. Note that when this 
cutoff time is reached, all un-decayed resonances are allowed to 
decay with their natural lifetimes and their projected decay 
positions and times are recorded. The validity of the numerical
methods used in the rescattering code have recently been studied using
the subdivision method, the results of which have verified that the methods used are 
valid \cite{Humanic:2006b}.

In carrying out a two-pion HBT study the two-pion coincident count rate 
is calculated along with the
one-pion count rates for reference. The two-pion
correlation function for pions binned in momenta $\mathbf{p_1}$ and
$\mathbf{p_2}$, $C(\mathbf{p_1},\mathbf{p_2})$, is constructed event-by-event from
the coincident countrate, $N_2(\mathbf{p_1},\mathbf{p_2})$ and
one-pion countrate summed over events, $N_1(\mathbf{p})$, as
\begin{equation}
\label{correxp}
C(\mathbf{p_1},\mathbf{p_2})=\frac{N_2(\mathbf{p_1},
\mathbf{p_2})}{N_1(\mathbf{p_1})N_1(\mathbf{p_2})}.
\end{equation}
It is usually
convenient to express the six-dimensional
$C(\mathbf{p_1},\mathbf{p_2})$ in terms of the four-vector momentum
difference, $Q=|p_1-p_2|$ by summing Eq.(\ref{correxp}) over
momentum difference,
\begin{eqnarray}
\label{correxp2} C(Q)=\sum_{\mathbf{p_1},\mathbf{p_2}(Q)}
\frac{N_2(\mathbf{p_1},
\mathbf{p_2})}{N_1(\mathbf{p_1})N_1(\mathbf{p_2})}=
\frac{A(Q)}{B(Q)},
\end{eqnarray}
where $A(Q)$ represents the ``real'' coincident two-pion countrate and
$B(Q)$ the ``background'' two-pion countrate composed of products
of the one-pion countrates,
all expressed in $Q$.~\cite{Lisa:2005a} In practice, $B(Q)$ is the
mixed event distribution, which is computed by forming pion-pairs
from different events.

For the HBT calculations from the model, a three-dimesional 
two-pion correlation function is formed using Eq.(\ref{correxp2})
and a Gaussian function in momentum difference variables is fitted to it to 
extract the pion source
parameters similar to what is done in experiments\cite{Humanic:2006a,Lisa:2005a}.  
Boson statistics are introduced after the
rescattering has finished
using the standard method of pair-wise symmetrization of bosons in
a plane-wave approximation \cite{Humanic:1986a}. The three-dimensional
correlation function, $C(Q_{side},Q_{out},Q_{long})$, is then calculated 
in terms of the momentum-difference
variables $Q_{side}$, which points in
the direction of the sum of the two pion momenta in the transverse
plane, $Q_{out}$, which points perpendicular to $Q_{side}$ in the
transverse plane and the longitudinal variable along the beam
direction $Q_{long}$.

The final step in the calculation is extracting fit parameters by
fitting a Gaussian parameterization to the model-generated two-pion correlation 
function given by, \cite{Lisa:2005a}
\begin{eqnarray}
\label{e6}
\lefteqn{C(Q_{side},Q_{out},Q_{long}) = } \\ &
G[ 1 + \lambda \exp( - Q_{side}^{2}R_{side}^{2} -
Q_{out}^{2}R_{out}^{2} - Q_{long}^{2}R_{long}^{2} ] \nonumber 
\end{eqnarray}
where the $R$-parameters, called the radius parameters, are associated with each 
momentum-difference variable direction, G is a normalization constant, and
$\lambda$ is the usual empirical parameter added to help in the
fitting of Eq. (\ref{e6}) to the actual correlation function
($\lambda = 1$ in the ideal case). The fit is carried out in the conventional LCMS 
frame (longitudinally comoving system) in which the
longitudinal pion pair momentum vanishes \cite{Lisa:2005a}.

\begin{figure}
\begin{center}
\includegraphics[width=80mm]{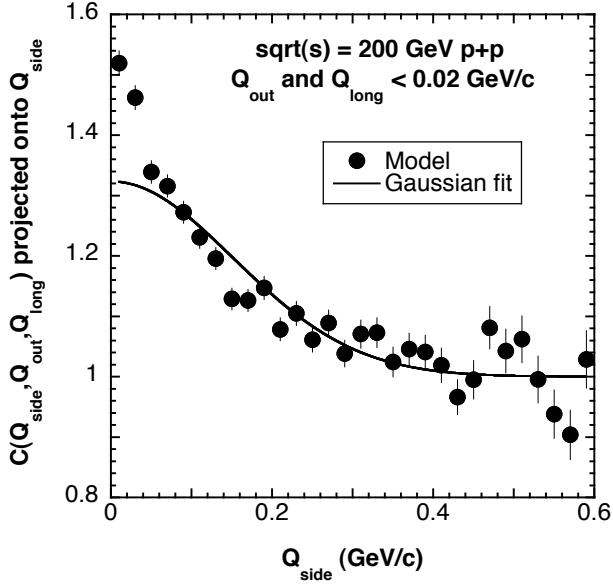} \caption{Sample two-pion correlation
function with three-dimensional Gaussian fit (Eq.(\ref{e6})) projected onto the $Q_{side}$
axis from the Model. The other variables, $Q_{out}$ and $Q_{long}$, are integrated
up to 0.02 GeV/c.
The cuts on the pion momenta are
$-0.5<y<0.5$, $0.15<p_T<0.8$ GeV/c, and $0.15<k_T<0.25$ GeV/c.}
\label{fig1}
\end{center}
\end{figure}

Figure \ref{fig1} shows a sample three-dimensional correlation function from the model
projected onto the $Q_{side}$ axis with projected fit
to Eq. (\ref{e6}). The other variables, $Q_{out}$ and $Q_{long}$, are integrated
up to 0.02 GeV/c. The cuts on the pion momenta are
$-0.5<y<0.5$, where y is rapidity, $0.15<p_T<0.8$ GeV/c, where $p_T$ is transverse
momentum, and $0.15<k_T<0.25$ GeV/c (see below). 
These cuts in $y$ and $p_T$ are used throughout to duplicate those
used in experiments. Although only
a small fraction of the full correlation function is shown in this plot, it gives some idea of
the quality of the Gaussian fit to the model. For $Q_{side}<0.05$ GeV/c
the fit is seen to undershoot the model which is a common feature also seen in experiments for
$p+p$ collisions\cite{Alexopoulos:1992iv,Bailly:1988zb}. The $\lambda$ parameter for this
fit is 0.330, which is far less than the ideal value of unity, another feature commonly seen in
$p+p$ experiments. The source of this non-Gaussian and non-ideal behavior in the model
is the presence of medium to long-lived resonances such as the $\omega$, $\eta$, and $\eta$'
which give a component to the correlation function representing a much larger pion source than
the majority of the pions in the collision and thus producing the narrower shape in the correlation function near $Q=0$.

\begin{figure}
\begin{center}
\includegraphics[width=85mm]{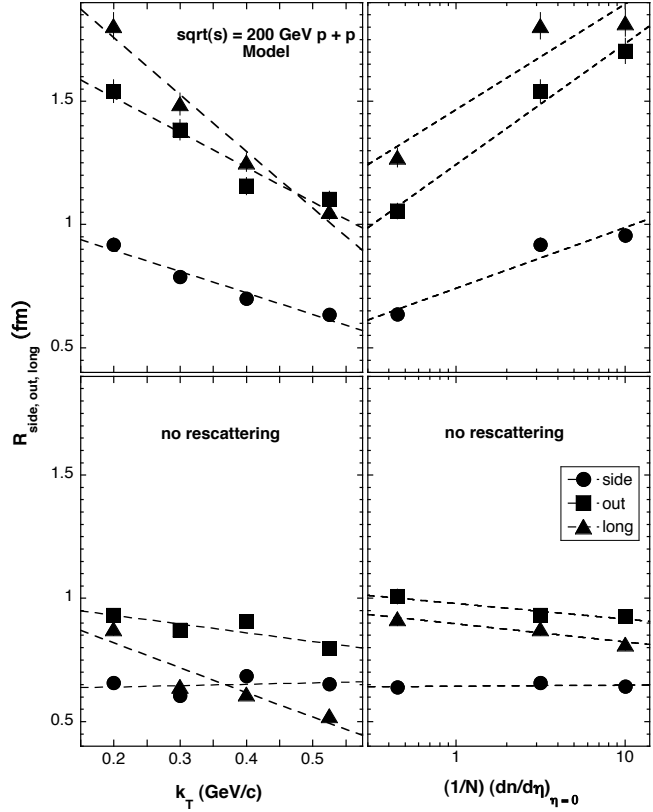} \caption{$k_T$ and rapidity density
dependences of the radius parameters from the model. The top plots show the
full model calculations and the bottom plots show model 
calculations in which hadronic rescattering is turned off. The dashed lines are
linear and logarithmic fits to the model points for the $k_T$ and rapidity density
plots, respectively.}
\label{fig2}
\end{center}
\end{figure}

Figure \ref{fig2} shows the pion momenta and particle multiplicity dependences of the radius
parameters from the model, both for the full model calculations (top plots)  and for model 
calculations in which hadronic rescattering is turned off (bottom plots). The pion momenta
and particle multiplicity are represented by the quantities 
$\mathbf{k_T}=(\mathbf{p_{T1}}+\mathbf{p_{T2}})/2$ which is the average transverse 
pion momentum
of the pair, and $(1/N) dn/d\eta$ evaluated at $\eta=0$ , which is the (pseudo)rapidity density
of charged particles at mid-rapidity, respectively. The dashed lines are fits 
to the model points, linear for
the $k_T$-dependence plots and logarithmic for the $(1/N) dn/d\eta$-dependence plots.
Focusing on the top two plots first, the full model calculations, it is clear that the
radius parameters have the same qualitative dependences on $k_T$ and rapidity density
as observed in heavy-ion collisions, namely decreasing with increasing $k_T$ and
increasing with increasing rapidity density (i.e. particle multiplicity). Another trend seen
in Figure \ref{fig2} which is also observed in heavy-ion collisions is
$R_{long}>R_{out}>R_{side}$. That hadronic rescattering is the main source of 
these effects in the model is seen by comparing the top plots with the bottom plots for which
rescattering is turned off in the model. In addition to the overall scales of the radius parameters
being significantly smaller with rescattering turned off, it is also seen that all of the dependences
which were seen in the top plots are either greatly reduced or absent in the
bottom plots. A small degree of the $k_T$ dependence is seen to linger for 
$R_{out}$ and $R_{long}$
which is mainly caused by the resonances present in the model.

\begin{figure}
\begin{center}
\includegraphics[width=75mm]{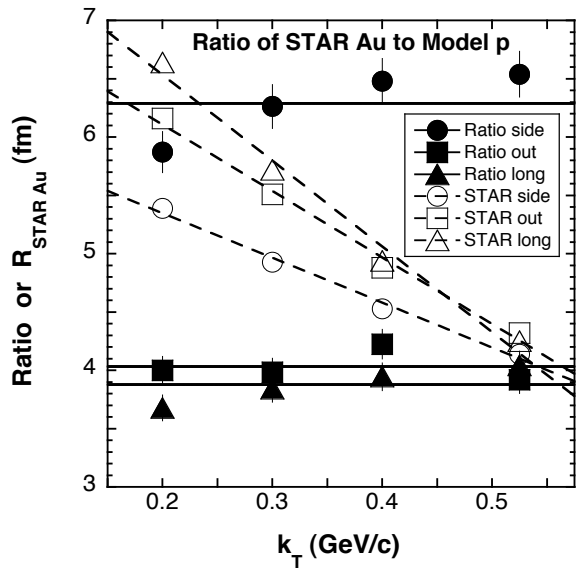} \caption{Ratios of STAR $Au+Au$
radius parameters to model $p+p$ radius parameters vs. $k_T$. The
STAR radius parameters used in calculating the ratios are also plotted.
The horizontal solid lines are the averages over $k_T$ of the ratios for
a given radius parameter. The dashed lines are linear fits to the STAR
radius parameters.}
\label{fig3}
\end{center}
\end{figure}

In order to make a more quantitative comparison 
between the $k_T$ dependence seen in Figure \ref{fig2}
in the full model for $p+p$ collisions and that measured in heavy-ion collisions, 
a calculation has been made of the ratios of the radius parameters extracted in the STAR experiment
for $Au+Au$ collisions at $\sqrt{s_{NN}}=200$ GeV\cite{Adams:2004yc} to those in Figure \ref{fig2}.
The same kinematic cuts on the pions as used in Figure \ref{fig2} were used by the STAR experiment
and the STAR results were extracted for a centrality cut of 0-5\%. Figure \ref{fig3} shows a
plot of these ratios vs. $k_T$. The STAR radius parameters used in calculating the
ratios are also plotted. Comparing the $k_T$ dependence of the model $p+p$ in the upper
left plot of Figure \ref{fig2} with the STAR radius parameters shown in Figure \ref{fig3}
the qualitative similarity of the plots is evident.
As seen, the ratios are approximately flat in $k_T$, but with a slight
increasing tendency hinting that the decrease of the $p+p$ radius parameters with $k_T$
from the model is slightly stronger than that measured in the $Au+Au$ data. Preliminary
measurements have been made of radius parameters from $p+p$ collisions at
$\sqrt s=200$ GeV by STAR and experimental ratios have been calculated as
in Figure \ref{fig3} showing a similarly flat dependence on $k_T$ \cite{Chajecki:2005zw}.

It is difficult to make a quantitative comparison between the model $p+p$ results for the
rapidity density shown in Figure \ref{fig2} and heavy-ion experiments in a similar way
as was done for the $k_T$ dependence since the range in rapidity density needed
to represent the $Au+Au$ measurements is about 100-650, which is much larger
than that seen in Figure \ref{fig2} of 0.3-10. A qualitative comparison can be made
in that it can easily be shown that the rapidity density dependence of STAR $Au+Au$
radius parameters from $\sqrt{s_{NN}}=200$ GeV collisions
is approximately logarithmically increasing with increasing rapidity density as is seen
in Figure \ref{fig2} \cite{Adams:2004yc,Back:2002wb}.

In conclusion, it has been shown that flow-like effects observed in relativistic
heavy-ion collision experiments can be reproduced in $p+p$ collisions 
at $\sqrt s=200$ GeV by a simple model
based on relativistic geometry and final-state hadronic rescattering with a short
proper time for hadronization of 0.1 fm/c. In the model, the flow-like effects are driven by
the hadronic rescattering which in turn is sensitively controlled by the hadronization time which
sets the initial particle density. If a long hadronization time were 
used, e.g. $\tau=1$ fm/c, the initial particle
density would be low and little rescattering would take place such that the full model
results would more resemble the bottom plots in Figure \ref{fig2}. Thus to the extent
of the agreement shown above between the present model calculations and experiment, another
implicit result of this study is that the hadronization time in $p+p$ collisions at this
energy appears to be very short.

\begin{acknowledgments}
The author wishes to acknowledge financial support from the U.S.
National Science Foundation under grant PHY-0653432, and to acknowledge computing
support from the Ohio Supercomputing Center.
\end{acknowledgments}

\end{document}